\documentclass[prd,aps,showpacs,secnumarabic]{revtex4}
\usepackage[dvips]{graphicx,color}
\usepackage{amssymb,amsmath,bm}
\def\ep{\epsilon}

\newcommand{\ts}{\thinspace}

\begin{document}
\title{An analytic description of the damping of gravitational waves by free streaming neutrinos.}
\author{Ben A. Stefanek}\email{stefanek@wisc.edu}
\author{Wayne W. Repko}\email{repko@pa.msu.edu} \affiliation{Department of Physics and Astronomy, Michigan State University, East Lansing, MI 48824}

\date{\today}

\begin{abstract}
We provide an analytic solution to the general wavelength integro-differential equation describing the damping of tensor modes of gravitational waves 
due to free streaming neutrinos in the early universe. Our result is expressed as a series of spherical Bessel functions whose coefficients are functions 
of the reduced wave number $Q$.
\end{abstract}
\maketitle

\section{Introduction}
Observations of the cosmic microwave background (CMB) have given increasing support to inflationary cosmological models. Density perturbations during 
this inflationary period are believed to have given rise to the large scale structures of the universe \cite{GHS}. 
In addition to these scalar perturbations, a spectrum of gravitational waves is also produced \cite{SP} (tensor perturbations) which could provide 
information about the early universe. In particular, the contribution of these tensor modes (measured in terms of a tensor-to-scalar ratio \cite{K}) 
to the temperature anisotropy of the CMB and to B-mode polarization of CMB photons could be used to check the predictions of inflationary models. Here we shall consider the effect of anisotropic 
inertia on the gravitational radiation and confirm that it is non-negligible. Although it will be assumed here that the anisotropic inertia is dominated by 
neutrinos and antineutrinos \cite{SW}, Ref. \cite{WZ} has proposed a free streaming relativistic gas contribution and a method to constrain its fraction
density through measurements of the CMB B-polarization spectra.

In a 2004 paper \cite{SW}, Weinberg derived an integro-differential equation for the propagation of cosmological gravitational waves. He writes a wave equation for the perturbation to the metric $h_{ij}({\bf x},t)$ and defines $\chi(u)$ as
\begin{equation}
h_{ij}=h_{ij}(0) \ts \chi(u)\,,
\end{equation}
where $u$ is the conformal time multiplied by the wave number
\begin{equation}
u=k \int^{t} \frac{dt'}{a(t')}\,.
\end{equation}
The wave equation for the perturbation leads to an integro-differential equation which the function $\chi$ satisfies for general wavelengths. In the variable $y=a(t)/a_{EQ}$, where $a_{EQ}$ is the expansion parameter at matter-radiation equality, this is Eq.\,(32) of \cite{SW}
\begin{equation}
(1+y)\chi''(y) + \left( \frac{2(1+y)}{y} + \frac{1}{2}\right) \chi'(y) 
+ Q^{2}\chi(y) = -\frac{24 \ts f_{\nu}(0)}{y^{2}} \int_{0}^{y} K(y,y') \frac{d\chi(y')}{dy'} dy'\,,
\label{eq:yEQ}
\end{equation}
with the initial conditions:
\begin{equation}
\chi(0)=1\,, \hspace{10mm} \chi'(0)=0\,.
\label{eq:yInitCond}
\end{equation}
Here $f_{\nu}(0)=0.40523$ is the fraction of the energy density in neutrinos and $Q=\sqrt{2}k/k_{EQ}$. The kernel $K$ will be discussed below, and $y$ is related to $u$ in the following manner
\begin{equation}
u=2Q\left( \sqrt{1+y}-1 \right)\equiv Q\,s\,.
\end{equation}

We will provide an analytic solution of Eqs.\,(\ref{eq:yEQ}) and (\ref{eq:yInitCond}) that is valid for general wavelengths. It will be shown that the effect of the anisotropic inertia is to damp the amplitude of the 
perturbation relative to the case where free-streaming neutrinos are absent, and that, in general, this damping depends on $Q$. Thus, it is important to have a solution capable of describing the damping for all wavelengths. 
For $Q^2\gg 1$, it has been shown that the solution for $\chi(u)$ can be written as a series of spherical Bessel functions whose coefficients are independent of $Q$ \cite{DR}. 
In \cite{SW}, Weinberg analyzed the $Q^2\gg 1$ limit, provided results for the damping factor when $Q^2\ll 1$ and made some observations about the damping for general value of $Q$. Our aim here is to show that it 
is possible to extend the spherical Bessel function expansion \cite{DR,GS} for $\chi(u)$ to the case of general $Q$. Having done this, we can recapture the $Q^2\ll 1$ results of \cite{SW}, the $Q^2\gg 1$ results 
of \cite{DR} and obtain precise results for intermediate values of $Q$. 

In the next section, we derive the equation that must be satisfied by a spherical Bessel function expansion for $\chi(u)$, examine its form in 
the $Q^2\gg 1$, obtain a recurrence relation for general $Q$, and use this relation to determine the damping factor for $Q\ll 1$. Section 3 contains a discussion of general wavelength case and we end with some conclusions. Lists of the various expansion coefficients and the details related to the solution of the recurrence relation are contained in the Appendices.
\section{Solution of the General Wavelength Equation}
In terms of the variable $u$, Eq.\,(\ref{eq:yEQ}) becomes
\begin{equation}
\chi''(u) + \frac{4(u+2Q)}{u^{2}+4Qu}\chi'(u) + \chi(u) = -\frac{16CQ^{2}}{(u^{2}+4Qu)^{2}}
\int_{0}^{u} K(u-u') \ts \chi'(u') \ts du'\,,
\label{eq:theDiffEQ}
\end{equation}
with $C = 24f_{\nu}(0) = 9.72552$ and the initial conditions
\begin{equation}
\chi(0)=1\,, \hspace{10mm} \chi'(0)=0\,.
\end{equation}
The kernel $K$ is a linear combination of spherical Bessel functions
\begin{equation}
K(z) = \frac{1}{16}\int_{-1}^{1} dx \ts (1-x^{2})^{2} e^{ixz}
=-\frac{\sin(z)}{z^{3}}-\frac{3\cos(z)}{z^{4}}+\frac{3\sin(z)}{z^{5}}
= \frac{1}{15}j_{0}(z)+\frac{2}{21}j_{2}(z)+\frac{1}{35}j_{4}(z)\,,
\end{equation}
which suggests a solution to Eq.\,(\ref{eq:theDiffEQ}) exists in the form of a series of spherical Bessel functions. Indeed the work done in Ref.\,\cite{DR} on the short wavelength limit shows that the convolution of a series of spherical Bessel functions with the kernel returns another series of spherical Bessel functions, so we look for a solution of the form
\begin{equation}
\chi(u) = \sum_{n=0}^{\infty} \alpha_{n}(Q) \ts j_{n}(u)\,.
\label{eq:besselModel}
\end{equation}

Multiplying Eq.\,(\ref{eq:theDiffEQ}) through by $(u^{2}+4Qu)^{2}$, expanding all the terms, and dividing by $u^{4}$ yields
\begin{multline}
\left(1+\frac{8Q}{u}+\frac{16Q^{2}}{u^{2}}\right)\chi''(u) + \left(\frac{4}{u}+\frac{24Q}{u^{2}}+\frac{32Q^{2}}{u^{3}}\right)\chi'(u) \\
+\left(1+\frac{8Q}{u}+\frac{16Q^{2}}{u^{2}}\right)\chi(u) = -\frac{16CQ^{2}}{u^{4}}
\int_{0}^{u} K(u-u') \ts \chi'(u') \ts du'\,.
\label{eq:expEq}
\end{multline}
The action of the differential operator on $j_{n}(u)$ gives:
\begin{multline}
\left[\left(1+\frac{8Q}{u}+\frac{16Q^{2}}{u^{2}}\right)\frac{d^{2}}{du^{2}} + \left(\frac{4}{u}+\frac{24Q}{u^{2}}+\frac{32Q^{2}}{u^{3}}\right)\frac{d}{du}
+\left(1+\frac{8Q}{u}+\frac{16Q^{2}}{u^{2}}\right)\right]j_{n}(u)\\
=\frac{1}{u^{4}}\left[n(n+1)16Q^{2} + n(n+2)8Qu + n(n+3)u^{2} \right]j_{n}(u)
-\frac{1}{u^{4}}\left[ 2u^{2}(u+4Q) \right]j_{n+1}(u)\,,
\end{multline}

and Eq.\,(\ref{eq:expEq}) becomes
\begin{multline}
\sum_{n=0}^{\infty} \alpha_{n}(Q) \left[n(n+1)16Q^{2} + n(n+2)8Qu + n(n+3)u^{2} \right]j_{n}(u)
-2\sum_{n=0}^{\infty} \alpha_{n}(Q)(u^{3}+4Qu^{2})j_{n+1}(u) \\
=-16CQ^{2}\sum_{n=0}^{\infty} \alpha_{n}\int_{0}^{u} K(u-u') \ts \frac{d\hspace{0.1mm}j_{n}(u')}{du'} \ts du'\,.
\end{multline}
If we divide through by the highest power, the Bessel function recurrence relation (AS 10.2.18) \cite{AS} 
\begin{equation}
\frac{j_{n}(z)}{z} = \frac{j_{n-1}(z)+j_{n+1}(z)}{(2n+1)}
\label{eq:JoverZ}
\end{equation} 
can be used to eliminate the powers of $u$. Dividing through by $u^3$ yields
\begin{multline}
\sum_{n=0}^{\infty} \alpha_{n}(Q) \left[\frac{16Q^{2}}{u^{3}}n(n+1) + \frac{8Q}{u^{2}}n(n+2) + \frac{1}{u}n(n+3) \right]j_{n}(u)
-2\sum_{n=0}^{\infty} \alpha_{n}(Q)\left(1+\frac{4Q}{u}\right)j_{n+1}(u) \\
=-\frac{16CQ^{2}}{u^{3}}\sum_{n=0}^{\infty} \alpha_{n}(Q)\int_{0}^{u} K(u-u') \ts \frac{d\hspace{0.1mm}j_{n}(u')}{du'} \ts du'\,.
\label{eq:genEQ}
\end{multline}
Notice that Eq.\,(\ref{eq:genEQ}) involves the same convolution integral for any value of $Q$. The limits $Q\ll 1$ and $Q\gg 1$ are of particular interest, but a recurrence relation for the $\alpha_n(Q)$ can be obtained for any $Q$. The simplest situation occurs for $Q\gg 1$ since, in this limit, the $\alpha_n(Q)$ reduce to numerical coefficients independent of $Q$ and the left hand side of Eq.\,(\ref{eq:genEQ}) reduces to a single term. 
\subsection{Short Wavelength Limit}
In the short wavelength limit $Q \gg 1$ the terms with $Q^{2}$ dominate and Eq.\,(\ref{eq:genEQ}) becomes \cite{AR}
\begin{equation}
\sum_{n=0}^{\infty} \alpha_{n} \ts n(n+1) j_{n}(u)
=-C\sum_{n=0}^{\infty} \alpha_{n}\int_{0}^{u} K(u-u') \ts \frac{d\hspace{0.1mm}j_{n}(u')}{du'} \ts du'\,.
\label{eq:e1}
\end{equation}
The derivative of $j_{n}(z)$ is given by (AS 10.2.19)\cite{AS}
\begin{equation} \label{eq:derj}
\frac{d\hspace{0.1mm}j_{n}(z)}{dz} = \frac{nj_{n-1}(z)-(n+1)j_{n+1}(z)}{(2n+1)}\,,
\end{equation}
and Eq.\,(\ref{eq:e1}) becomes
\begin{equation}
\sum_{n=0}^{\infty} \alpha_{n} \ts n(n+1) j_{n}(u)
=-C\sum_{n=0}^{\infty} \frac{\alpha_{n}}{(2n+1)}\int_{0}^{u} du' K(u-u') \ts \left[ nj_{n-1}(u')-(n+1)j_{n+1}(u')\right]\,.
\end{equation}
We can shift indices on the Bessel functions if we define $\alpha_{n}=0$ for $n<0$ to obtain
\begin{equation}
\sum_{n=0}^{\infty} \alpha_{n} \ts n(n+1) j_{n}(u)
=-C\sum_{n=0}^{\infty} \omega_{n}\int_{0}^{u} du' \ts K(u-u') \ts j_{n}(u')\,,
\label{eq:e2}
\end{equation}
with
\begin{equation}
\omega_{n} = \left[\frac{(n+1)\,\alpha_{n+1}}{(2n+3)}-\frac{n\ts\alpha_{n-1}}{(2n-1)}\right]\,.
\end{equation}
The convolution integral can be evaluated using the technique given in Ref.\,\cite{DR}
\begin{equation}
\sum_{n=0}^{\infty} \omega_{n}\int_{0}^{u} du' \ts K(u-u') \ts j_{n}(u') = \sum_{n=0}^{\infty} \epsilon_{n} \ts j_{n}(u)\,,
\label{eq:RD}
\end{equation}
where
\begin{equation}\label{epsn}
\epsilon_{n} = \frac{(2n+1)}{2} \ts i^{n} \left( \sum_{\ell=0}^{\infty} \sum_{m=0,2,4} d_{m} \ts \omega_{\ell} \ts (-i)^{\ell+m+1} \ts I_{n \ell}^{m} \right)\,,
\end{equation}
and
\begin{equation}\label{In}
I_{n \ell}^{m} = \int_{-1}^{1} ds \ts P_{n}(s) \left[ Q_{m}(s)P_{\ell}(s) + P_{m}(s)Q_{\ell}(s) \right]\,.
\end{equation}
Here, $P_n(s)$ and $Q_n(s)$ are the Legendre polynomial and the Legendre function of the $2^{\rm nd}$ kind, and
\begin{equation}\label{dm}
d_{0} = \frac{1}{15}, \thickspace d_{2} = \frac{2}{21}, \thickspace d_{4} = \frac{1}{35}\,.
\end{equation}
Then, Eq. (\ref{eq:e2}) becomes
\begin{equation}
\sum_{n=0}^{\infty} \alpha_{n} \ts n(n+1) j_{n}(u)
=-C\sum_{n=0}^{\infty} \epsilon_{n} \ts j_{n}(u)\,.
\end{equation}
This yields the following recurrence relation
\begin{equation}
\alpha_{n} \ts n(n+1)=-C\epsilon_{n}\,.
\label{eq:rec0}
\end{equation}
The $\epsilon_{n}$ can be found using Eqs.\,(\ref{epsn}), (\ref{In}) and (\ref{dm}). The first 20 are given in Appendix A by taking $Q\gg 1$. For $n=0$, Eq.\, (\ref{eq:rec0}) vanishes, so $\alpha_{0}$ is undetermined. However, the initial condition $\chi(0)=1$ fixes it to unity.
The next six equations are
\begin{eqnarray*}
2 \alpha_1 &=& -\frac{C \alpha _ 1}{15} \\
6 \alpha_2 &=& -C \left(\frac{\alpha _ 2}{15}-\frac{\alpha _ 0}{6}\right) \\
12 \alpha_3 &=& -C \left(\frac{\alpha _ 3}{15}-\frac{\alpha _ 1}{18}\right) \\
20 \alpha_4 &=& -C \left(-\frac{\alpha _ 0}{10}-\frac{\alpha _ 2}{20}+\frac{\alpha _ 4}{15}\right) \\
30 \alpha_5 &=& -C \left(-\frac{11 \alpha _1}{225}-\frac{33 \alpha _3}{700}+\frac{\alpha _5}{15}\right)\\
42 \alpha_6 &=& -C \left(-\frac{13 \alpha _0}{900}-\frac{13 \alpha _2}{315}-\frac{143 \alpha _4}{3150}+\frac{\alpha _6}{15}\right)\\
\end{eqnarray*}
The first equation requires $\alpha_{1}=0$, and we see that all the odd terms depend recursively on $\alpha_{1}$. Thus, all the odd terms vanish in the large $Q$ limit. Solving for the first 4 non-zero $\alpha_{n}$ gives
\begin{equation}
\alpha_0=1.0\,,\quad \alpha_{2} = 0.243807\,, \quad \alpha_{4} = 5.28424 \times 10^{-2}\, \quad \alpha_{6} = 6.13545 \times 10^{-3} \,,
%\alpha_{8} &=& 2.97534 \times 10^{-4} \\
%\alpha_{10} &=& 6.16273 \times 10^{-5} \\
\end{equation}
in agreement with Ref.\,\cite{DR}. For large $Q$, the full solution $\chi(s,Q)$ is
\begin{equation}
\chi(s,Q)\sim \sum_{n=0}^\infty\,(-1)^n\alpha_{2n}\frac{\sin(Qs)}{Qs}\,,
\end{equation}
while the undamped solution $\chi_0(s,Q)$ is $j_0(Qs)=\sin(Qs)/Qs$. The damping factor is therefore
\begin{equation}
\left|\frac{\chi(s,Q)}{\chi_0(s,Q)}\right|^2=\left|\sum_{n=0}^\infty\,(-1)^n\alpha_{2n}\right|^2\,.
\end{equation}
\subsection{Recurrence Relation}
Returning to the general case given by Eq. (\ref{eq:genEQ}), we can apply Eq.\,(\ref{eq:RD}) to the righthand side and obtain
\begin{multline}
\sum_{n=0}^{\infty} \alpha_{n} \left[\frac{16Q^{2}}{u^{3}}n(n+1) + \frac{8Q}{u^{2}}n(n+2) + \frac{1}{u}n(n+3) \right]j_{n}(u)
-2\sum_{n=0}^{\infty} \alpha_{n}\left(1+\frac{4Q}{u}\right)j_{n+1}(u) \\
=-\frac{16CQ^{2}}{u^{3}}\sum_{n=0}^{\infty} \epsilon_{n} \ts j_{n}(u)\,.
\label{eq:genEQ2}
\end{multline}
Eq.\,(\ref{eq:JoverZ}) can then be used recursively to obtain (see Appendix B)
\begin{equation}
\sum_{n=0}^{\infty} \left[ 16Q^{2}\beta_{n}+8Q(\gamma_{n}-\theta_{n})+\delta_{n}-2\alpha_{n-1}\right] j_{n}(u)
=-16CQ^{2}\sum_{n=0}^{\infty} \lambda_{n} \ts j_{n}(u)\,.
\label{eq:rec4}
\end{equation}
where $\beta_{n},\gamma_{n},\delta_{n},\theta_{n},\lambda_{n}$ are defined in Appendix B and presented in Appendix C. Eq.\,(\ref{eq:rec4}) leads to the recurrence relation
\begin{equation}
16Q^{2}(\beta_{n}+C\lambda_{n})+8Q(\gamma_{n}-\theta_{n})+\delta_{n}-2\alpha_{n-1}=0\,.
\label{eq:RecRel}
\end{equation}

The equations given by Eq.\,(\ref{eq:RecRel}) determine the coefficients of the spherical Bessel functions in Eq.\,(\ref{eq:besselModel}). With these coefficients, Eq.\,(\ref{eq:besselModel}) provides an analytic solution to the inhomogeneous Eq.\,(\ref{eq:theDiffEQ}) for general values of the wave number $Q$. Eq.\,(\ref{eq:RecRel}) can also provide the coefficients of the homogeneous solution (the solution without free streaming neutrinos) by setting $C=0$, yielding the following recurrence relation
\begin{equation}
16Q^{2}\beta_{n}+8Q(\gamma_{n}-\theta_{n})+\delta_{n}-2\alpha_{n-1}=0\,.
\label{eq:homoRecRel}
\end{equation}

The coefficients were found by solving Eqs.\,(\ref{eq:RecRel}) and (\ref{eq:homoRecRel}) using \emph{Mathematica} (with the condition $\alpha_{n}=0$ for $n<0$). The first 20 inhomogeneous and homogeneous coefficients are given in Appendix A. The $a_{n}$ are the inhomogeneous coefficients and the $b_{n}$ are the homogeneous coefficients. This convention will be used throughout 
the remainder of the paper.

\subsection{The Long Wavelength Limit}
We found an analytic solution to Eq.\,(\ref{eq:theDiffEQ}) consisting of an expansion in spherical Bessel functions of the following form
\begin{equation}
\chi(s,Q) = \sum_{n=0}^{\infty} \alpha_{n}(Q) \ts j_{n}(Q\,s)\,,
\label{eq:sBesselModel}
\end{equation}
where the $\alpha_{n}(Q)$ are explicit functions of the wave number $Q$. Upon first glance at the terms in the Appendix A, this expansion seems to be divergent in the limit $Q \rightarrow 0$ since the coefficients depend inversely on $Q$. However, the expansion is in fact finite due to the implicit presence of $Q$ in $j_{n}(u)$. Since both the full solution $\chi(u)$ and the homogeneous solution $\chi_0(u)$ are equal to $1$ as $Q\to 0$, the comparison of the full and homogeneous solutions should be made using the ratio of $\chi'(s,Q)$ to $\chi'_0(s,Q)$. Using Eq.\,(\ref{eq:derj}) and $u=Qs$, $\chi'(u)$ can be expressed as
\begin{equation}
\chi'(s,Q)=\sum_{n=0}^\infty a_n(Q)\left(\frac{nj_{n-1}(Qs)-(n+1)j_{n+1}(Qs)}{2n+1}\right)Q\,.
\end{equation}
To find the first few terms in the low $Q$ expansion, we use
\begin{equation}
j_n(x)\to  \frac{x^n \sqrt{\pi}} {2^{1+n}\Gamma\left(3/2+n\right)}\left(1-\frac{x^2}{2^2\left(3/2+n\right)} +\frac{x^4} {2^4\left(5/2+n\right)\left(3/2+n\right)}+O[x^6]\right)\,.
\end{equation}
The cosmologically interesting value of $s$ is the value at last scattering, where $y=22.1\Omega_Mh^2$. Using $\Omega_Mh^2=0.15$, $s_L=2.15452$. With $x=2.15454\,Q$ and the $a_n(Q)$ in Appendix A, the low $Q$ result is
\begin{equation}\label{chi'low}
\chi'(s_{L},Q) = -0.573661 \ts Q^2 + 0.243294 \ts Q^4 -0.0381643 \ts Q^6 +\mathcal{O}(Q^8)\,.
\end{equation}
Using the same method, we obtain the following expansion of the homogeneous term 
\begin{equation}\label{chi0'low}
\chi'_{0}(s_{L},Q) = -0.601254 \ts Q^2 + 0.264482 \ts Q^4 - 0.0424186 \ts Q^6 +\mathcal{O}(Q^8)\,.
\end{equation}
The $a_n(Q)$ and $b_n(Q)$ in Appendix A are sufficient to determine the numerical coefficients in Eqs.\,(\ref{chi'low}) and (\ref{chi0'low}) to one part in $10^5$. Weinberg damping factor $|\chi'(s_{L},Q)/\chi'_{0}(s_{L},Q)|^{2}$ for $Q=0$ can be computed as
\begin{equation}
\left|\frac{\chi'(s_{L},0)}{\chi'_{0}(s_{L},0)}\right|^{2} = \lim_{Q \to 0} \left|\frac{-0.573661 \ts Q^2}{-0.601254 \ts Q^2}\right|^{2} = 0.91032\,,
\end{equation}
in agreement with Weinberg's discussion on the low $Q$ limit. As Weinberg remarks ``This damping is relatively insensitive to $Q$ for small $Q$." 
Confirmation of this statement is seen here where the damping is independent of $Q$ until the fourth order terms become important. The finiteness of 
this damping factor shows that the solution is well defined for arbitrarily small values of $Q$.

\section{Damping for General Wavelengths}
\begin{figure}[h!]
\centering
\includegraphics[scale=0.75]{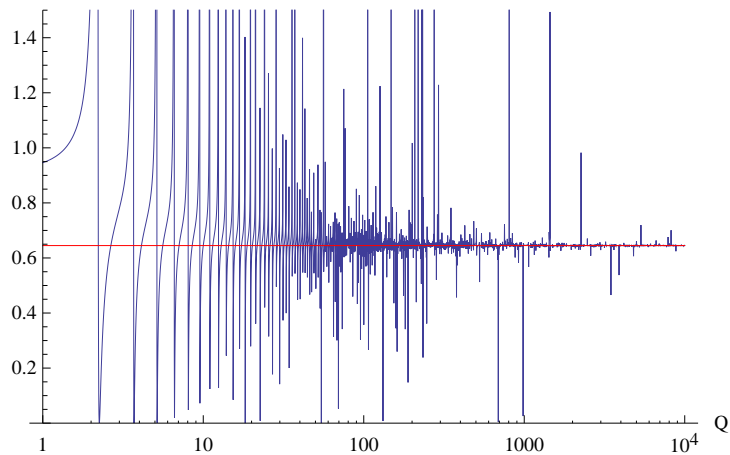}
\caption{The damping factor $|\chi'(s_{L},Q)/\chi'_{0}(s_{L},Q)|^{2}$ is shown along with the horizontal line representing $A^{2}=0.645019$.}
\label{fig:weinDamp_genQ}
\end{figure}

For $Q>1$, there is no difficulty with the coefficients $a_n(Q)$ (or $b_n(Q)$) and asymptotically this expansion reproduces the damping found in Ref.\,  \cite{DR} for $Q\gg 1$, as shown in Figure\,\ref{fig:weinDamp_genQ}. It should be noted that the plotted damping factor is the ratio of two oscillating functions and therefore diverges at the zeros of the function in the denominator. However, since the functions are only slightly out of phase, these spikes occur where $\chi'(s_{L},Q)$ is small and according to Ref.\,\cite{SW}, this makes the spikes uninteresting since the multipole coefficients for the corresponding values of $\ell$ will be very difficult to measure.

For the most part, 20 terms were used in the numerical evaluations of Eq.\,(\ref{eq:sBesselModel}) occurring in this paper, although 100 terms where generated for comparison purposes. The observation in Ref.\, \cite{SW} that the reduction in $|\chi'(s_{L},Q)/\chi'_{0}(s_{L},Q)|^{2}$ 
from $1$ for $Q=0.55$ and $Q=0.8$ is about 8\% and 7\% respectively is confirmed in Table\,\ref{Qdamps2}. Also shown in this Table is the trend of $|\chi'(s_{L},Q)/\chi'_{0}(s_{L},Q)|^{2}$ in the relatively flat regions 
between the spikes steadily decrease from the value $\approx$ 0.9 for Q $\ll$ 1 to a value close to 0.644 for Q $\approx$ 10. 

\begin{table}[h!]
\centering
\begin{tabular}{| c | c | c |}
   \hline
   $Q$ & $|\chi(s_{L},Q)/\chi_{0}(s_{L},Q)|^{2}$ & $|\chi'(s_{L},Q)/\chi'_{0}(s_{L},Q)|^{2}$ \\
   \hline
  0.01 & 1.00001 & 0.910275 \\
   \hline
  0.1 & 1.00081 & 0.910561 \\
   \hline
  0.55 & 1.02745 & 0.919509 \\
   \hline
  0.8 & 1.06802 & 0.931291 \\
   \hline
  1 & 1.13058 & 0.946036 \\
   \hline
  10 & 0.811905 & 0.644909 \\
   \hline
  $10^{2}$ & 0.64902 & 0.866294 \\
   \hline
  $10^{3}$ & 0.646315 & 0.642628 \\
   \hline
  $10^{4}$ & 0.644227 & 0.64506 \\
   \hline
  $10^{5}$ & 0.645004 & 0.645055 \\
   \hline
  $10^{6}$ & 0.645013 & 0.645008 \\
   \hline
\end{tabular}
\caption{Damping values for various $Q$'s from Eq.\,(\ref{eq:sBesselModel})}
\label{Qdamps2}
\end{table}

\section{Conclusions}
We have shown that the treatment of gravitational wave damping by free streaming neutrinos can be framed in terms of a series of spherical Bessel functions for all values of the reduced wave number $Q$. The result for the coefficients of the spherical Bessel series when $Q \gg 1$ \cite{DR} emerges quite simply from the coefficient recurrence relation for a general $Q$. For the opposite limit, $Q \ll 1$, the analysis can then be extended to arbitrarily small values of $Q$ by using the low $Q$ expansion of $j_n(Q\,s_L)$. For intermediate values of $Q$, the coefficients obtained from the general recurrence relation can be used directly for any $Q$, provided that the factor $Q$ is retained in the argument of the spherical Bessel functions.  

There have been numerous successful numerical studies of the damping of the gravitational wave spectrum due to anisotropic inertia (Refs. \cite{WK}, \cite{Kuro}, \cite{Latt}, \cite{Kin}). The present approach represents a trade off between the use of reliable sophisticated numerical techniques to calculate $\chi(s,Q)$ and the necessity of evaluating the exact series solution for $\chi(s,Q)$ to sufficient accuracy. The main advantage of using the series expansion is that the convolution of the desired solution with the kernel can be evaluated exactly in terms of spherical Bessel functions. 

\begin{figure}[h!]
\centering
\includegraphics[scale=0.75]{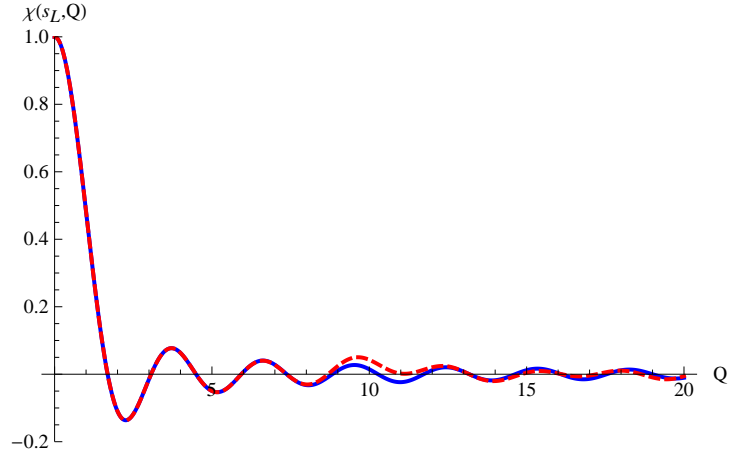}
\caption{The comparison between $\chi(s_L,Q)$ with 20 terms (red dashed) and $\chi(s_L,Q)$ with 100 terms (blue) is shown for $Q\leq 20$.}
\label{diff}
\end{figure}

We used 20 terms in our expansions of $\chi(s_L,Q)$ and $\chi_0(s_L,Q)$. To illustrate the accuracy that this truncated series provides, Fig.\,\ref{diff}, compares the amplitude $\chi(s_L,Q)$ computed with 20 terms and 100 terms.
As another test of the use of 20 terms rather than 100 terms in the series Eq.\,(\ref{eq:sBesselModel}), we compare the ratio of the damping factor $R=|\chi'(s_L,Q)/\chi'_0(s_L,Q)|^2$ for the two cases as a function of $Q$  in Table\,\ref{Delta}.
\begin{table}[h!]
\centering
\begin{tabular}{| c | c |}
\hline
$Q$        & $R_{100}/R_{20}$ \\
\hline
  0.01     & $1.00003$  \\
   \hline
  0.1      & $1.00003$  \\
   \hline
  1        & $1.00002$  \\
   \hline
  10       & $1.04926$  \\
   \hline
  $10^{2}$ & $1.37376$  \\
   \hline
  $10^{3}$ & $0.99772$  \\
   \hline
  $10^{4}$ & $1.00005$  \\
   \hline
  $10^{5}$ & $1.00002$  \\
   \hline
  $10^{6}$ & $0.99999$  \\
   \hline
\end{tabular}
\caption{The ratio of the damping factor $|\chi'(s_L,Q)/\chi'_0(s_L,Q)|^2$ is shown for 20 and 100 terms in the spherical Bessel function expansion.}
\label{Delta}
\end{table}
The only region where the reduction in the number of terms significantly affects the answer is the transition region $10\leq Q\leq 100$. If this 
region is important, additional terms would be required, but are readily computable from the provided recurrence relations.

\begin{acknowledgements}
The authors would like to thank Duane Dicus for useful comments and WWR acknowledges numerous discussions with Gavriil Shchedrin. This work was supported in part by the National Science Foundation under Grant PHY 1068020.
\end{acknowledgements}

\appendix
\section{Expansion Coefficients}
The relations between the $\ep_n(Q)$ and the expansion coefficients $a_n(Q)$ are listed below for $n=0$ to $14$.
\begin{eqnarray*}
\ep_0(Q) &=& 0 \\ [2pt]
\ep_1(Q) &=& \frac{a_1(Q)}{15} \\
\ep_2(Q) &=& \frac{a_2(Q)}{15}-\frac{a_0(Q)}{6} \\
\ep_3(Q) &=& \frac{a_3(Q)}{15}-\frac{a_1(Q)}{18} \\
\ep_4(Q) &=& \frac{a_4(Q)}{15}-\frac{a_2(Q)}{20}-\frac{a_0(Q)}{10} \\
\ep_5(Q) &=& \frac{a_5(Q)}{15}-\frac{33a_3(Q)}{700}-\frac{11a_1(Q)}{225} \\
\ep_6(Q) &=& \frac{a_6(Q)}{15}-\frac{143a_4(Q)}{3150}-\frac{13a_2(Q)}{315} -\frac{13a_0(Q)}{900} \\
\ep_7(Q) &=& \frac{a_7(Q)}{15}-\frac{13a_5(Q)}{294}-\frac{11a_3(Q)}{294} -\frac{5a_1(Q)}{588} \\
\ep_8(Q) &=& \frac{a_8(Q)}{15}-\frac{17a_6(Q)}{392}-\frac{221a_4(Q)}{6300} -\frac{17a_2(Q)}{2520}+\frac{17a_0(Q)}{11025} \\
\ep_9(Q) &=& \frac{a_9(Q)}{15}-\frac{323a_7(Q)}{7560}-\frac{19a_5(Q)}{567} -\frac{4693a_3(Q)}{793800}+\frac{19a_1(Q)}{18900} \\
\ep_{10}(Q) &=& \frac{a_{10}(Q)}{15}-\frac{19a_8(Q)}{450}-\frac{17a_6(Q)}{525} -\frac{41a_4(Q)}{7560}+\frac{a_2(Q)}{1260}-\frac{a_0(Q)}{2520} \\
\ep_{11}(Q) &=& \frac{a_{11}(Q)}{15}-\frac{23a_9(Q)}{550} -\frac{437a_7(Q)}{13860} -\frac{69989a_5(Q)}{13721400} +\frac{299a_3(Q)}{436590}-\frac{23a_1(Q)}{83160} \\
\ep_{12}(Q) &=& \frac{a_{12}(Q)}{15}-\frac{115a_{10}(Q)}{2772}-\frac{5a_8(Q)}{162} -\frac{18715a_6(Q)}{3841992}+\frac{85a_4(Q)}{137214}-\frac{a_2(Q)}{4536} +\frac{25a_0(Q)}{174636} \\
\ep_{13}(Q) &=& \frac{a_{13}(Q)}{15}-\frac{15a_{11}(Q)}{364}-\frac{69 a_9(Q)}{2275}-\frac{887 a_7(Q)}{188760}+\frac{13319a_5(Q)}{23123100}  
-\frac{1751a_3(Q)}{9249240}+\frac{a_1(Q)}{9555} \\
\ep_{14}(Q) &=& \frac{a_{14}(Q)}{15}-\frac{261
a_{12}(Q)}{6370}-\frac{145a_{10}(Q)}{4851}-\frac{2850091 a_8(Q)}{624323700}+\frac{8207a_6(Q)}{15057900}-\frac{551 a_4(Q)}{3243240} 
+\frac{29a_2(Q)}{343035}-\frac{29a_0(Q)}{463320} \\
\ep_{15}(Q) &=& \frac{a _{15}(Q)}{15}-\frac{899 a _{13}(Q)}{22050}-\frac{31
a _{11}(Q)}{1050}-\frac{589 a _9(Q)}{132132}+\frac{713 a _7(Q)}{1366365}-\frac{2077 a _5(Q)}{13250952}+\frac{589 a _3(Q)}{8108100}-\frac{31 a_1(Q)}{653400} \\
\ep_{16}(Q) &=& \frac{a_{16}(Q)}{15}-\frac{341a_{14}(Q)}{8400} -\frac{319a_{12}(Q)}{10920} -\frac{12247a_{10}(Q)}{2802800} +\frac{59a_8(Q)}{117117}-\frac{23 a_6(Q)}{156156}+\frac{163a_4(Q)}{2509650} -\frac{a_2(Q)}{25740}+ \frac{a_0(Q)}{32175} \\
\ep_{17}(Q) &=& \frac{a _{17}(Q)}{15}-\frac{11 a _{15}(Q)}{272}-\frac{31 a _{13}(Q)}{1071}-\frac{319 a_{11}(Q)}{74256}+\frac{1681 a _9(Q)} {3435432}-\frac{197 a _7(Q)}{1405404}+\frac{1679 a _5(Q)}{28158273}-\frac{343 a _3(Q)}{10239372}+\frac{35 a_1(Q)}{1444014} \\
\ep_{18}(Q) &=& \frac{a_{18}(Q)}{15}-\frac{37 a_{16}(Q)}{918}-\frac{407 a _{14}(Q)}{14175}-\frac{42439a_{12}(Q)}{10024560}+\frac{3992633a_{10}(Q)}{8362153800} -\frac{925a_8(Q)}{6870864}+\frac{37 a_6(Q)}{663663}-\frac{851 a_4(Q)}{28442700} \\
&&+\frac{37a _2(Q)}{1840410}-\frac{37a _0(Q)}{2187900}   \\
\ep_{19}(Q) &=& \frac{a_{19}(Q)}{15}-\frac{481a_{17}(Q)}{11970}-\frac{13a_{15}(Q)}{456} -\frac{714467a_{13}(Q)}{170931600}+\frac{31a_{11}(Q)}{66300} -\frac{29a_9(Q)}{222768}+\frac{346a_7(Q)}{6551545}-\frac{167 a_5(Q)}{6096948}+\frac{4646a_3(Q)}{266741475} \\
&&-\frac{a_1(Q)}{74100} \\
\ep_{20}(Q) &=& \frac{a_{20}(Q)}{15}-\frac{533
a_{18}(Q)}{13300}-\frac{1517a_{16}(Q)}{53550}-\frac{8651a_{14}(Q)}{2093040} +\frac{153053a_{12}(Q)}{333316620}-\frac{47027a_{10}(Q)}{371651280} +\frac{72529a_8(Q)}{1435925400} \\
&&-\frac{101147a_6(Q)}{3963016200} +\frac{6478 a_4(Q)}{416116701} 
-\frac{41a_2(Q)}{3619980}+\frac{41a_0(Q)}{4157010}     
\end{eqnarray*}

Using $C=24f_\nu(0)=9.72552$, the first twenty $a_n(Q)$ are
\begin{eqnarray*}
a_0(Q) &=& 1 \\ [6pt]
a_1(Q) &=& 0 \\ [6pt]
a_2(Q) &=& 0.243807 \\ [4pt]
a_3(Q) &=& \frac{0.843856}{Q} \\
a_4(Q) &=& 0.0528424 -\frac{1.31658}{Q^2} \\
a_5(Q) &=& \frac{2.60385}{Q^3}-\frac{1.98354}{Q} \\
a_6(Q) &=& 0.00613545 -\frac{6.16442}{Q^4}+\frac{8.15545}{Q^2} \\
a_7(Q) &=& \frac{16.7669}{Q^5}-\frac{31.6633}{Q^3}+\frac{2.81654}{Q} \\
a_8(Q) &=& 0.000297534 -\frac{50.564}{Q^6}+\frac{130.451}{Q^4} -\frac{22.4418}{Q^2} \\
a_9(Q) &=& \frac{162.551}{Q^7}-\frac{583.689}{Q^5}+\frac{141.989}{Q^3} -\frac{3.66375}{Q} \\
a_{10}(Q) &=& 0.0000616273 -\frac{524.266}{Q^8}+\frac{2845.09}{Q^6} -\frac{868.128}{Q^4} +\frac{46.8401}{Q^2} \\
a_{11}(Q) &=& \frac{1458.97}{Q^9}-\frac{15066.7}{Q^7}+\frac{5429.37}{Q^5} -\frac{433.843}{Q^3}+\frac{4.49219}{Q} \\
a_{12}(Q) &=&-4.998662\times 10^{-6}-\frac{1012.45}{Q^{10}} +\frac{86302.4}{Q^8} -\frac{35476.}{Q^6}+\frac{3653.68}{Q^4} -\frac{83.7481}{Q^2} \\
a_{13}(Q) &=& -\frac{39324.1}{Q^{11}}-\frac{532143.}{Q^9}+\frac{244196.}{Q^7} 
-\frac{30143.3}{Q^5}+\frac{1064.03}{Q^3} -\frac{5.3138}{Q} \\
a_{14}(Q) &=& 2.33661\times 10^{-6}+\frac{570180.}{Q^{12}}+\frac{3.5156\times 10^6}{Q^{10}}-\frac{1.7758\times 10^6}{Q^8}+\frac{251558.}{Q^6}  
-\frac{11768.6}{Q^4}+\frac{135.604}{Q^2}\\
a_{15}(Q)&=&-\frac{6.31661\times 10^6}{Q^{13}}-\frac{2.47754\times 10^7}{Q^{11}}+\frac{1.36468\times 10^7}{Q^9}-\frac{2.15679\times 10^6}{Q^7}+\frac{123426.}{Q^5}-\frac{2258.33}{Q^3}+\frac{6.13067}{Q} \\
a_{16}(Q) &=&-8.58528\times 10^{-7}+\frac{6.56547\times 10^7}{Q^{14}}+\frac{1.85489\times 10^8}{Q^{12}}-\frac{1.10726\times 10^8}{Q^{10}}+\frac{1.91474\times 10^7}{Q^8}-\frac{1.27662\times 10^6}{Q^6} \\ &&+\frac{31691.5}{Q^4}-\frac{204.83}{Q^2} \\
a_{17}(Q) &=&-\frac{6.776\times 10^8}{Q^{15}}-\frac{1.46977\times 10^9}{Q^{13}}+\frac{9.47105\times 10^8}{Q^{11}}-\frac{1.76711\times 10^8}{Q^9}+\frac{1.32929\times 10^7}{Q^7}-\frac{411129.}{Q^5}\\ &&+\frac{4323.5}{Q^3}-\frac{6.94444}{Q} \\
a_{18}(Q) &=&3.76663\times 10^{-7}+\frac{7.10223\times 10^9}{Q^{16}}+\frac{1.22823\times 10^{10}}{Q^{14}}-\frac{8.52514\times 10^9}{Q^{12}}+\frac{1.69845\times 10^9}{Q^{10}}-\frac{1.40956\times 10^8}{Q^8} \\
&&+\frac{5.154\times 10^6}{Q^6}-\frac{74973.1}{Q^4}+\frac{293.848}{Q^2} \\
a_{19}(Q) &=& -\frac{7.63956\times 10^{10}}{Q^{17}}-\frac{1.07891\times 10^{11}}{Q^{15}}+\frac{8.06027\times 10^{10}}{Q^{13}}-\frac{1.70115\times 10^{10}}{Q^{11}}+\frac{1.53236\times 10^9}{Q^9}-\frac{6.39674\times 10^7}{Q^7} \\
&&+\frac{1.17764\times 10^6}{Q^5}-\frac{7659.4}{Q^3}+\frac{7.75604}{Q} \\
a_{20}(Q) &=& -1.79691\times 10^{-7}+\frac{8.47695\times 10^{11}}{Q^{18}}+\frac{9.9312\times 10^{11}}{Q^{16}}-\frac{7.98966\times
10^{11}}{Q^{14}}+\frac{1.77533\times 10^{11}}{Q^{12}}-\frac{1.71459\times 10^{10}}{Q^{10}} \\
&&+\frac{7.97294\times 10^8}{Q^8}-\frac{1.75791\times 10^7}{Q^6} +\frac{160721.}{Q^4} -\frac{405.077}{Q^2}
\end{eqnarray*}
For large $Q$, the coefficients of the $a_{2n}(Q)$ reduce to the results in Ref.\,\cite{DR}. For a fixed value of $y$, these coefficients and the $b_n(Q)$ below yield finite results.

The coefficients for the homogeneous solution are
\begin{eqnarray*}
b_0(Q) &=& 1 \\ [6pt]
b_1(Q) &=& 0 \\ [6pt]
b_2(Q) &=& 0 \\ [6pt]
b_3(Q) &=& \frac{1.45833}{Q} \\
b_4(Q) &=& -\frac{2.95313}{Q^2} \\
b_5(Q) &=& \frac{7.57969}{Q^3}-\frac{2.56667}{Q} \\
b_6(Q) &=& -\frac{23.4609}{Q^4}+\frac{12.4583}{Q^2} \\
b_7(Q) &=& \frac{84.8364}{Q^5}-\frac{55.7289}{Q^3}+\frac{3.61607}{Q} \\
b_8(Q) &=& -\frac{350.539}{Q^6}+\frac{261.588}{Q^4}-\frac{31.5725}{Q^2} \\
b_9(Q) &=& \frac{1628.06}{Q^7}-\frac{1323.89}{Q^5}+\frac{218.214}{Q^3} -\frac{4.64444}{Q} \\
b_{10}(Q) &=& -\frac{8391.91}{Q^8}+\frac{7261.36}{Q^6}-\frac{1451.12}{Q^4} +\frac{63.3263}{Q^2} \\
b_{11}(Q) &=& \frac{47522.4}{Q^9}-\frac{43105.3}{Q^7}+\frac{9831.73}{Q^5} -\frac{624.673}{Q^3}+\frac{5.66288}{Q} \\
b_{12}(Q) &=& -\frac{293207.}{Q^{10}}+\frac{275999.}{Q^8}-\frac{69352.2}{Q^6} +\frac{5591.11}{Q^4}-\frac{110.739}{Q^2}\\
b_{13}(Q) &=& \frac{1.9574\times 10^6}{Q^{11}}-\frac{1.89854\times 10^6}{Q^9}+\frac{513799.}{Q^7}-\frac{48915.7}{Q^5}+\frac{1475.69}{Q^3} -\frac{6.67582}{Q} \\
b_{14}(Q) &=& -\frac{1.4056\times 10^7}{Q^{12}}+\frac{1.39746\times 10^7}{Q^{10}}-\frac{4.01079\times 10^6}{Q^8}+\frac{431978.}{Q^6} -\frac{17104.2}{Q^4}+\frac{176.825}{Q^2} \\
b_{15}(Q) &=& \frac{1.08026\times 10^8}{Q^{13}}-\frac{1.09654\times 10^8}{Q^{11}}+\frac{3.30103\times 10^7}{Q^9}-\frac{3.91143\times 10^6}{Q^7}+\frac{187766.}{Q^5}-\frac{3057.83}{Q^3}+\frac{7.68542}{Q} \\
b_{16}(Q) &=& -\frac{8.84662\times 10^8}{Q^{14}}+\frac{9.14024\times 10^8}{Q^{12}}-\frac{2.86269\times 10^8}{Q^{10}}+\frac{3.66061\times 10^7}{Q^8}-\frac{2.03032\times 10^6}{Q^6}+\frac{44547.7}{Q^4} -\frac{264.595}{Q^2} \\
b_{17}(Q) &=& \frac{7.6902\times 10^9}{Q^{15}}-\frac{8.06762\times 10^9}{Q^{13}}+\frac{2.61251\times 10^9}{Q^{11}}-\frac{3.55547\times 10^8}{Q^9}+\frac{2.20733\times 10^7}{Q^7}-\frac{599623.}{Q^5} +\frac{5758.72}{Q^3}-\frac{8.69281}{Q} \\
b_{18}(Q) &=& -\frac{7.07184\times 10^{10}}{Q^{16}}+\frac{7.51825\times 10^{10}}{Q^{14}}-\frac{2.50508\times 10^{10}}{Q^{12}}+\frac{3.59103\times 10^9}{Q^{10}}-\frac{2.4409\times 10^8}{Q^8}+\frac{7.79394\times 10^6}{Q^6}-\frac{102980.}{Q^4} \\
&&+\frac{377.055}{Q^2} \\
b_{19}(Q) &=& \frac{6.85875\times 10^{11}}{Q^{17}}-\frac{7.37751\times 10^{11}}{Q^{15}}+\frac{2.51961\times 10^{11}}{Q^{13}}-\frac{3.77431\times 10^{10}}{Q^{11}}+\frac{2.76404\times 10^9}{Q^9}-\frac{1.00219\times 10^8}{Q^7} \\
&&+\frac{1.66778\times 10^6}{Q^5}-\frac{10082.}{Q^3}+\frac{9.69868}{Q}\\
b_{20}(Q) &=& -\frac{6.99674\times 10^{12}}{Q^{18}}+\frac{7.60439\times 10^{12}}{Q^{16}}-\frac{2.6537\times 10^{12}}{Q^{14}}+\frac{4.12814\times 10^{11}}{Q^{12}}-\frac{3.21801\times 10^{10}}{Q^{10}}+\frac{1.29312\times 10^9}{Q^8} \\ 
&&-\frac{2.56593\times 10^7}{Q^6}+\frac{217086.}{Q^4}-\frac{517.212}{Q^2}
\end{eqnarray*}

\section{Recurrence Calculation}
We want to eliminate powers of $u$ from
\begin{multline}
\sum_{n=0}^{\infty} \alpha_{n} \left[\frac{16Q^{2}}{u^{3}}n(n+1) + \frac{8Q}{u^{2}}n(n+2) + \frac{1}{u}n(n+3) \right]j_{n}(u)
-2\sum_{n=0}^{\infty} \alpha_{n}\left(1+\frac{4Q}{u}\right)j_{n+1}(u) \\
=-\frac{16CQ^{2}}{u^{3}}\sum_{n=0}^{\infty} \epsilon_{n} \ts j_{n}(u)
\label{eq:AgenEQ2}
\end{multline}

The Bessel function recurrence relation
\begin{equation}
\frac{j_{n}(z)}{z} = \frac{j_{n-1}(z)+j_{n+1}(z)}{(2n+1)}
\label{eq:AJoverZ}
\end{equation} 
can be used recursively to obtain the following relations: 
\begin{equation}
\frac{j_{n}(z)}{z^{2}} = \frac{j_{n-2}(z)}{(2n-1)(2n+1)}+\frac{2j_{n}(z)}{(2n-1)(2n+3)}+\frac{j_{n+2}(z)}{(2n+1)(2n+3)}
\label{eq:JoverZ2}
\end{equation}
\begin{equation}
\begin{split}
\frac{j_{n}(z)}{z^{3}} &= \frac{j_{n-3}(z)}{(2n+1)(2n-1)(2n-3)}+\frac{3j_{n-1}(z)}{(2n-3)(2n+1)(2n+3)}\\
 &+\frac{3j_{n+1}(z)}{(2n-1)(2n+1)(2n+5)}+\frac{j_{n+3}(z)}{(2n+1)(2n+3)(2n+5)}
\end{split}
\label{eq:JoverZ3}
\end{equation}

Consider the first term on the LHS of Eq. (\ref{eq:AgenEQ2})
\begin{multline}
\sum_{n=0}^{\infty} \alpha_{n} \left[\frac{16Q^{2}}{u^{3}}n(n+1) + \frac{8Q}{u^{2}}n(n+2) + \frac{1}{u}n(n+3) \right]j_{n}(u)\\
=16Q^{2}\sum_{n=0}^{\infty} \alpha_{n} \ts n(n+1) \ts \frac{j_{n}(u)}{u^{3}}+8Q\sum_{n=0}^{\infty} \alpha_{n} \ts n(n+2) \ts \frac{j_{n}(u)}{u^{2}}
+\sum_{n=0}^{\infty} \alpha_{n} \ts n(n+3) \ts \frac{j_{n}(u)}{u}
\label{eq:rec1}
\end{multline}

The first term in Eq. (\ref{eq:rec1}) can be re-written as follows with the aid of Eq. (\ref{eq:JoverZ3}):
\begin{multline}
16Q^{2}\sum_{n=0}^{\infty} \alpha_{n} \ts n(n+1) \ts \frac{j_{n}(u)}{u^{3}}
=16Q^{2}\sum_{n=0}^{\infty} \alpha_{n} \ts n(n+1) \bigg[ \frac{j_{n-3}(z)}{(2n+1)(2n-1)(2n-3)}
+\frac{3j_{n-1}(z)}{(2n-3)(2n+1)(2n+3)}\\
+\frac{3j_{n+1}(z)}{(2n-1)(2n+1)(2n+5)}+\frac{j_{n+3}(z)}{(2n+1)(2n+3)(2n+5)} \bigg]
\end{multline}

If we define $\alpha_{n}=0$ for $n<0$ then the indices on the Bessel functions can be shifted to obtain:
\begin{multline}
16Q^{2}\sum_{n=0}^{\infty} \alpha_{n} \ts n(n+1) \ts \frac{j_{n}(u)}{u^{3}}
=16Q^{2}\sum_{n=0}^{\infty} j_{n}(u) \bigg[ \frac{\alpha_{n+3} \ts (n+3)(n+4)}{(2n+3)(2n+5)(2n+7)}
+\frac{3 \ts \alpha_{n+1} \ts (n+1)(n+2)}{(2n-1)(2n+3)(2n+5)}\\
+\frac{3 \ts \alpha_{n-1} \ts n(n-1)}{(2n-3)(2n-1)(2n+3)}+\frac{\alpha_{n-3} \ts (n-3)(n-2)}{(2n-5)(2n-3)(2n-1)} \bigg]
\end{multline}

So we can write:
\begin{equation}
16Q^{2}\sum_{n=0}^{\infty} \alpha_{n} \ts n(n+1) \ts \frac{j_{n}(u)}{u^{3}}
=16Q^{2}\sum_{n=0}^{\infty} \beta_{n} \ts j_{n}(u)
\end{equation}
with:
\begin{equation}
\begin{split}
\beta_{n} &= \bigg[ \frac{\alpha_{n+3} \ts (n+3)(n+4)}{(2n+3)(2n+5)(2n+7)}
+\frac{3 \ts \alpha_{n+1} \ts (n+1)(n+2)}{(2n-1)(2n+3)(2n+5)}\\
&+\frac{3 \ts \alpha_{n-1} \ts n(n-1)}{(2n-3)(2n-1)(2n+3)}+\frac{\alpha_{n-3} \ts (n-3)(n-2)}{(2n-5)(2n-3)(2n-1)} \bigg]
\end{split}
\end{equation}

Following the same procedure for the rest of Eq. (\ref{eq:rec1}) using Eqs. (\ref{eq:AJoverZ}) and (\ref{eq:JoverZ2}) yields:
\begin{equation}
\sum_{n=0}^{\infty} \alpha_{n} \left[\frac{16Q^{2}}{u^{3}}n(n+1) + \frac{8Q}{u^{2}}n(n+2) + \frac{1}{u}n(n+3) \right]j_{n}(u)
=\sum_{n=0}^{\infty} j_{n}(u) \left[ 16Q^{2}\beta_{n}+8Q\gamma_{n}+\delta_{n} \right]
\label{eq:rec2}
\end{equation}
with:
\begin{equation}
\gamma_{n}=\left[\frac{\alpha_{n+2} \ts (n+2)(n+4)}{(2n+3)(2n+5)}+\frac{2 \ts \alpha_{n} \ts n(n+2)}{(2n-1)(2n+3)}
+\frac{\alpha_{n-2} \ts n(n-2)}{(2n-3)(2n-1)}\right]
\end{equation}
\begin{equation}
\delta_{n}=\left[\frac{\alpha_{n+1} \ts (n+1)(n+4)}{(2n+3)}+\frac{\alpha_{n-1} \ts (n-1)(n+2)}{(2n-1)}\right]
\end{equation}

Similarly, the second term on the LHS of Eq. (\ref{eq:AgenEQ2}) becomes:
\begin{equation}
-2\sum_{n=0}^{\infty} \alpha_{n}\left(1+\frac{4Q}{u}\right)j_{n+1}(u)
=-2\sum_{n=0}^{\infty} j_{n}(u) \left[4Q\theta_{n}+\alpha_{n-1}\right]
\end{equation}
with:
\begin{equation}
\theta_{n}=\left[\frac{\alpha_{n}}{(2n+3)}+\frac{\alpha_{n-2}}{(2n-1)}\right]
\end{equation}

So Eq. (\ref{eq:AgenEQ2}) becomes:
\begin{equation}
\sum_{n=0}^{\infty} \left[ 16Q^{2}\beta_{n}+8Q(\gamma_{n}-\theta_{n})+\delta_{n}-2\alpha_{n-1}\right] j_{n}(u)
=-\frac{16CQ^{2}}{u^{3}}\sum_{n=0}^{\infty} \epsilon_{n} \ts j_{n}(u)
\label{eq:rec3}
\end{equation}

With the condition $\alpha_{n}=0$ for $n<0$. The $\epsilon_{n}$ are already zero for $n<0$, so the same method as above can be applied to the RHS of
Eq. (\ref{eq:rec3}). Using Eq. (\ref{eq:JoverZ3}) gives:
\begin{multline}
-\frac{16CQ^{2}}{u^{3}}\sum_{n=0}^{\infty} \epsilon_{n} \ts j_{n}(u)
=-16CQ^{2}\sum_{n=0}^{\infty} j_{n}(u) \bigg[ \frac{\epsilon_{n+3}}{(2n+3)(2n+5)(2n+7)}
+\frac{3 \ts \epsilon_{n+1}}{(2n-1)(2n+3)(2n+5)}\\
+\frac{3 \ts \epsilon_{n-1}}{(2n-3)(2n-1)(2n+3)}+\frac{\epsilon_{n-3}}{(2n-5)(2n-3)(2n-1)} \bigg]
\end{multline}

So the RHS of Eq. (\ref{eq:rec3}) becomes:
\begin{equation}
-\frac{16CQ^{2}}{u^{3}}\sum_{n=0}^{\infty} \epsilon_{n} \ts j_{n}(u)
=-16CQ^{2}\sum_{n=0}^{\infty} \lambda_{n} \ts j_{n}(u)
\end{equation}
with:
\begin{equation}
\begin{split}
\lambda_{n}&=\bigg[ \frac{\epsilon_{n+3}}{(2n+3)(2n+5)(2n+7)}
+\frac{3 \ts \epsilon_{n+1}}{(2n-1)(2n+3)(2n+5)}\\
&+\frac{3 \ts \epsilon_{n-1}}{(2n-3)(2n-1)(2n+3)}+\frac{\epsilon_{n-3}}{(2n-5)(2n-3)(2n-1)} \bigg]
\end{split}
\end{equation}

Substituting, Eq. (\ref{eq:rec3}) becomes
\begin{equation}
\sum_{n=0}^{\infty} \left[ 16Q^{2}\beta_{n}+8Q(\gamma_{n}-\theta_{n})+\delta_{n}-2\alpha_{n-1}\right] j_{n}(u)
=-16CQ^{2}\sum_{n=0}^{\infty} \lambda_{n} \ts j_{n}(u)
\label{eq:Arec4}
\end{equation}
which is Eq. (\ref{eq:rec4}) of this paper.

\section{Recurrence Coefficients}

\begin{equation}
\begin{split}
\beta_{n} &= \bigg[ \frac{\alpha_{n+3} \ts (n+3)(n+4)}{(2n+3)(2n+5)(2n+7)}
+\frac{3 \ts \alpha_{n+1} \ts (n+1)(n+2)}{(2n-1)(2n+3)(2n+5)}\\
&+\frac{3 \ts \alpha_{n-1} \ts n(n-1)}{(2n-3)(2n-1)(2n+3)}+\frac{\alpha_{n-3} \ts (n-3)(n-2)}{(2n-5)(2n-3)(2n-1)} \bigg]
\end{split}
\end{equation}

\begin{equation}
\gamma_{n}=\left[\frac{\alpha_{n+2} \ts (n+2)(n+4)}{(2n+3)(2n+5)}+\frac{2 \ts \alpha_{n} \ts n(n+2)}{(2n-1)(2n+3)}
+\frac{\alpha_{n-2} \ts n(n-2)}{(2n-3)(2n-1)}\right]
\end{equation}

\begin{equation}
\delta_{n}=\left[\frac{\alpha_{n+1} \ts (n+1)(n+4)}{(2n+3)}+\frac{\alpha_{n-1} \ts (n-1)(n+2)}{(2n-1)}\right]
\end{equation}

\begin{equation}
\theta_{n}=\left[\frac{\alpha_{n}}{(2n+3)}+\frac{\alpha_{n-2}}{(2n-1)}\right]
\end{equation}

\begin{equation}
\begin{split}
\lambda_{n}&=\bigg[ \frac{\epsilon_{n+3}}{(2n+3)(2n+5)(2n+7)}
+\frac{3 \ts \epsilon_{n+1}}{(2n-1)(2n+3)(2n+5)}\\
&+\frac{3 \ts \epsilon_{n-1}}{(2n-3)(2n-1)(2n+3)}+\frac{\epsilon_{n-3}}{(2n-5)(2n-3)(2n-1)} \bigg]
\end{split}
\end{equation}

\end{document}